\documentclass[12pt,preprint]{aastex}

\def\ltsima{$\; \buildrel < \over \sim \;$}
\def\lsim{\lower.5ex\hbox{\ltsima}}
\def\gtsima{$\; \buildrel > \over \sim \;$}
\def\gsim{\lower.5ex\hbox{\gtsima}}

\received{2001 June 29}
\begin{document}

\title{Anomalous Star-Formation Activity of Less-Luminous Galaxies 
in
Cluster Environment} 

\author{Yuka Y. Tajiri\quad{\rm and}\quad Hideyuki
Kamaya\altaffilmark{1}}

\affil{Department of Astronomy, Kyoto University}
\affil{Sakyo-ku, Kyoto 606-8502, Japan}

\email{tajiri@kusastro.kyoto-u.ac.jp}
\and\email{kamaya@kusastro.kyoto-u.ac.jp}

\altaffiltext{1}{Visiting Academics at Department of Physics, 
Oxford
University, Keble Road, Oxford, OX1, 3RH, UK}

\begin{abstract}

We discuss a correlation between star formation activity (SFA) and
luminosity of star-forming galaxies at intermediate redshifts of 
$0.2\le
z\le 0.6$ in both cluster and field environments. Equivalent width 
(EW)
of [O{\sc ii}] is used for measurement of the SFA, and $R$-band 
absolute
magnitude, $M_R$, for the luminosity.  In less-luminous ($M_R\
\gsim -20.7$) galaxies, we find : (1) the mean EW([O{\sc ii}]) of
cluster galaxies is smaller than that of field galaxies; but (2) 
some
cluster galaxies have as large EW([O{\sc ii}]) as that of actively
star-forming field galaxies.  Based on both our results, we 
discuss a
new possible mechanism for the Butcher-Oemler (BO) effect, 
assuming that
the luminosity of a galaxy is proportional to its dynamical mass.  
Our
proposal is that BO galaxies are less-massive cluster galaxies 
with smaller
peculiar velocities. They are then stable against Kelvin-Helmholtz
instability (KHI), and are not affected by tidal interaction 
between
clusters and themselves.  Their interstellar medium (ISM) would be
hardly stripped, and their SFA would be little suppressed. Hence, 
as
long as such galaxies keep up their SFA, the fraction of blue 
galaxies
in a cluster does not decrease. As a cluster becomes virialized,
however, such galaxies become more accelerated, the ISM available 
for SFA is
stripped by KHI, and their color evolves redward, which produces 
the
BO effect.
\end{abstract}


\section{INTRODUCTION}

The Butcher-Oemler (BO) effect is the first observational evidence 
of the
evolution of cluster galaxy populations.  It shows that clusters 
at
higher redshift than 0.2 have a higher fraction of blue galaxies 
than
that of local clusters (Butcher \& Oemler 1978, 1984). In the last
decade, {\it Hubble Space Telescope (HST)} images have revealed 
that
such blue galaxies, which are recognized as BO galaxies, are
predominantly normal late-type spirals (e.g., Dressler {\it et 
al.}
1994; Couch {\it et al.}  1994, 1998).

According to numerical and analytical studies, a dynamical 
interaction
between a cluster and its galaxies has been proposed as a 
mechanism for
the BO effect. Because of this interaction, late-type spiral 
galaxies
lose and/or consume their interstellar medium (ISM) and then their 
star
formation activity (SFA) is suppressed during their infall into 
the
higher-density regions of a cluster. Their color evolves
redward, and the fraction of BO galaxies in the cluster
decreases. (e.g., Gunn \& Gott 1972; Sarazin 1986; Kodama \& Bower 
2001;
Abadi {\it et al.}  1999; Quilis {\it et al.}  2000; Mori \& 
Burkert
2000; Balogh {\it et al.}  2000). Indeed, many observations have
revealed that BO galaxies lie preferentially in the lower-density
regions of clusters \citep{Butcher1984, Dokkum2000, Balogh1999}.  
Thus,
observational results seem consistent with the current theoretical
picture.

The above studies predict that there should be a difference 
between
mean SFA of cluster galaxies and that of field galaxies at 
intermediate
redshifts of $0.2 \le z \le 0.6$. Furthermore, such a difference 
should
be more evident in less-luminous (i.e. less-massive) galaxies 
whose
gravitational potential wells are shallower and which are more
environmentally affected.  This trend, however, has not been
observationally confirmed since previous papers especially focused 
on
the morphology of galaxies at the redshifts.  In this $Letter$, we
present a statistical investigation of environmental effects on 
SFA for
less-luminous galaxies at intermediate redshifts.

In addition to the dynamical interaction that is mentioned above, 
other
effects for the BO phenomenon are also expected. The first one is
a decrease of the infall-rate of galaxies into clusters with time
(e.g., Kodama \& Bower 2001). As a second one, in this $Letter$ we
propose a further possible kinematic origin for the BO effect from 
our
new findings in observational data.

This {\it Letter} is organized as follows: In \S 2, we present the
results of our statistical analysis of a correlation between 
EW([O{\sc
ii}]) and luminosity of cluster and field galaxies at intermediate
redshifts. In \S 3, we discuss SFA in less-luminous cluster 
galaxies,
and propose an alternative cause for the BO effect. In \S 4, we 
give our conclusions. We assume a spatially flat universe with 
$\Omega
_0=0.3$, $\Lambda _0=0.7$ and $H_{0}=70$km/s/Mpc.

\section{ENVIRONMENTAL EFFECTS ON SFA}

Environmental effects on the SFA of galaxies can be statistically 
investigated
with archival data. As a measure of SFA, we adopt the equivalent 
width (EW)
of [O{\sc ii}] $\lambda$3727.  The [O{\sc ii}] emission is 
utilized
because it is correlated to Balmer-line emission that can trace 
current
SFA \citep{Gallagher1989}. EW is used because the aperture effect 
can be
neglected, and because it is line-luminosity normalized by 
luminosity of
whole a galaxy. This measure permits us to discuss the specific 
SFA of
each galaxy.

For cluster galaxies, those in the following papers are utilized : 
9
MORPHS clusters ($0.37<z<0.56$; $r\le22$; Dressler {\it et al.}  
1999);
Cl1358+62 ($z=0.33$; $r\le21$; Fisher {\it et al.}  1998); AC103, 
AC114,
and AC118 ($z=0.31$; $r\le22$; Couch {\it et al.}  1998); 
Abell2390
($z=0.23$; $r\le21.0$; Abraham {\it et al.}  1996); and CNOC
($0.18<z<0.56$; $M_R\le-19.6$; Balogh {\it et al.}  1999), where 
$M_R$
and $r$ are absolute magnitude and apparent magnitude in $R$-band,
respectively. For field galaxies, we use CNOC ($0.18<z<0.56$;
$M_R\le-19.6$; Balogh {\it et al.}  1999).  We compile a sample
of star-forming galaxies according to the following four 
conditions :
{\textbf{(i)}} galaxies are brighter than the limiting magnitude 
of each
catalog; {\textbf{(ii)}} galaxies are spectroscopically selected 
as
cluster or field galaxies according to \citet{Balogh1999};
{\textbf{(iii)}} the redshifts of field galaxies are within the 
range of
$0.2\le z\le 0.6$, which corresponds to that of all the clusters; 
and
{\textbf{(iv)}} EW([O{\sc ii}]) of galaxies is positive.

In Fig.1, we present EW([O{\sc ii}]) and $R$-band absolute 
magnitudes for the
galaxies. In panel (a), cluster galaxies are plotted, while field
galaxies are plotted in panel (b). With the same sample of Fig.1, 
we depict
the mean EW of galaxies in each magnitude bin in each environment, 
in
panels (a) and (b) of Fig.2. From these figures, we can examine
environmental effects on mean SFA. The results are more 
convincingly
revealed in Table 1 by means of Welch's $t$-test (Press {\it et 
al.}
1988). Even when variances are not the same between two 
sample-groups,
this test can be applied to answer the question of whether their 
averages 
are the same or not, using the $t$ value which is estimated as :
\begin{eqnarray}
t(M_{R_i})&=&\frac{{\rm EW}_{\rm cl}(M_{R_i})-{\rm EW}_{\rm 
f}(M_{R_i})}
{[{\rm Var}({\rm EW}_{\rm cl}(M_{R_i}))/N_{\rm cl}(M_{R_i})
+{\rm Var}({\rm EW}_{\rm f}(M_{R_i}))/N_{\rm f}(M_{R_i})]^{1/2}} .
\end{eqnarray}
Here, EW$_{\rm cl}(M_{R_i})$ is the equivalent width of cluster 
galaxies
in the $i$-th magnitude bin, EW$_{\rm f}(M_{R_i})$ that of field 
galaxies,
Var(EW) the variance of each value, $N_{\rm cl}(M_{R_i})$ the 
number of
cluster galaxies in the $i$-th magnitude bin, and $N_{\rm 
f}(M_{R_i})$ that
of field galaxies. In Table 1, we also give a non-parametric test,
Mann-Whitney's $U$-test \citep{Pitman1948}, to answer whether the
statistically representative values of two sample-groups are same 
or
not.  The $U$-test is performed since our sample may not be 
normally
distributed because of selection bias, {\textbf {(iv)}}. Table 1
contains the results of the $t$-test and $U$-test for all galaxies 
in 
the sample and also for galaxies with EW smaller than $100\AA$.
The suffix of $'i'$ runs from the bins of more-luminous galaxies 
to
those of less-luminous galaxies, and its maximum is adopted to be
eight\footnote{Since the number of bins generally affects 
statistical tests, 
we have checked that the general trend of the results is
not altered if the maximum bin number is 7, 9, or 10.}.

Adopting a significance level of 0.05, we first
conclude that {\textbf{(i)}} for less luminous galaxies with
magnitudes of $M_{R} \gsim -20.7$, the mean SFA of cluster 
galaxies is always
smaller than that of field galaxies. Even when we omit several 
galaxies
whose EW is larger than 100\AA, our conclusion is not altered. On 
the
assumption that luminosity is proportional to dynamical mass of a
galaxy, this result statistically confirms the expectation of
almost all theoretical studies that the SFA of less-luminous 
cluster
galaxies should suffer more suppression because of their shallower
gravitational potential well (e.g., Henriksen \& Byrd 1996; Abadi 
{\it
et al.}  1999; Quilis {\it et al.} 2000; Mori \& Burkert 2000; 
Balogh
{\it et al.}  2000).

As a second conclusion, while there is no theoretical prediction 
that
less-luminous cluster galaxies keep on their SFA, we discover the
following: {\textbf{(ii)}} there are some less-luminous cluster 
galaxies
with exceptionally large SFA (EW$\ge30$\AA: i.e. larger than the 
mean EW
in less-luminous, $M_R \gsim -20.7$, field galaxies), which is as 
large
as that of less-luminous but actively star-forming field galaxies. 
 The
existence of such cluster galaxies is consistent with the 
observational
fact that BO galaxies are mostly less luminous
\citep{Dressler1994}. Therefore, to argue the possibility of such
exceptions, we re-examine the dynamical effects on their ISM
in the next section.

\section{DISCUSSION} 

As long as less-luminous cluster galaxies are assumed to be less
massive, their ISM should be easily stripped by environmental 
effects,
and then their SFA must be suppressed. In this section, we 
inversely
examine which environmental effect is inactive in stripping the 
ISM in
less-luminous cluster galaxies. The typical effects are the 
following :
{\textbf{(i)}} tidal interaction between a cluster and its 
galaxies;
{\textbf{(ii)}} ram pressure owing to intracluster medium (ICM) ; 
and
{\textbf{(iii)}} Kelvin-Helmholtz instability (KHI) via 
disturbances on
the boundary of the ISM and ICM (e.g., Henriksen \& Byrd 1996). 
Since we
focus on less-luminous galaxies, we assume that luminosity of a 
galaxy
is proportional to its dynamical mass, and adopt the density 
profile of
Burkert \citep{Burkert1995}, which well explains that of 
less-massive
galaxies.  In that profile, the radius of a galaxy ($R_{\rm gal}$) 
is
related to its dynamical mass (${\cal M}_{\rm gal}$) by
$R_{\rm gal}=6.9\times10^{-4}({\cal M}_{\rm gal})^{3/7}$.

We firstly consider the tidal effect between a cluster and its 
galaxy,
comparing with the self-gravity of its galaxy.  The tidal force
($F_{\rm tid}$) and the gravitational force ($F_{\rm grav}$) are 
given
as
$F_{\rm tid}=\displaystyle{2G{\cal M}_{\rm cl}R_{\rm gal}/R_{\rm
cl}^3}$
and 
$F_{\rm grav}=\displaystyle{G{\cal M}_{\rm gal}/R_{\rm gal}^2},$
respectively. Here $G$ is the gravitational constant, ${\cal 
M}_{\rm
cl}$ the mass of a galaxy cluster, and $R_{\rm cl}$ the radius of 
a
galaxy cluster.  From the equilibrium of the two forces, we obtain 
a
critical mass :
\begin{eqnarray}
{\cal M}_{\rm tid}
&=&3\times10^{11}({R_{\rm cl}}/{500{\rm kpc}})^{21/2}({{\cal 
M}_{\rm cl}}
/{10^{14}M_{\odot}})^{7/2}M_{\odot}.
\label{eq:mass_tid}
\end{eqnarray}
This means that the tidal force is not effective for less-luminous
galaxies when their mass is smaller than ${\cal M}_{\rm tid}$, 
although
it does not prove anything for massive galaxies which the Burkert
profile may not apply.

Secondly, we take the ram pressure ($P_{\rm ram}$) due to the ICM, 
and
compare with the gravity $(P_{\rm grav})$ of a galaxy in the same
dimension of the pressure. These are expressed as
$P_{\rm ram}=\rho_{\rm ICM}\sigma_{\rm gal}^2$,
and 
$P_{\rm grav}=2\pi G\Sigma_{\rm gal}\Sigma_{\rm 
ISM}=\displaystyle{2G{\cal M}_{\rm
gal}^2F/\pi R_{\rm gal}^4}$,
respectively. Here $\rho_{\rm ICM}$ is the mass density of ICM,
$\sigma_{\rm gal}$ the velocity of a galaxy, $\Sigma_{\rm gal}$ 
the
surface density of a galaxy, $\Sigma_{\rm ISM}$ that of ISM, and 
$F$ the
ratio of gas mass to dynamical mass. Therefore, less-massive
(i.e. less-luminous) galaxies than a critical mass of:
\begin{eqnarray}
 {\cal M}_{\rm ram}
 &=&1\times10^7(F/0.1)^{-7/2}(\rho_{\rm ICM}/10^{-4}m_{\rm
 H})^{7/2}(\sigma_{\rm gal}/500{\rm km\, s^{-1}})^7M_{\odot}
\label{eq:mass_ram}
\end{eqnarray}
are affected by ram-pressure stripping effects.

Thirdly, we take up the KHI that is ineffective in a condition of:
\begin{eqnarray}
k=\frac{\pi}{R_{\rm gal}}<-\frac{G{\cal M}_{\rm gal}}{R_{\rm 
gal}^2} \,
\, \frac{\rho_{\rm ICM}^2 -\rho_{\rm ISM}^2}{\rho_{\rm 
ICM}\rho_{\rm
ISM}{\sigma_{\rm gal}^2}} \sim \frac{G{\cal M}_{\rm gal}^2 F}{\pi 
R_{\rm
gal}^5 \rho_{\rm ICM} \sigma_{\rm gal}^2},
\end{eqnarray}
where $k$ is the wave number \citep{Chand1961}, and $\rho_{\rm 
ISM}$ the
mass density of ISM. Thus, we find more-massive galaxies than a 
critical
mass of
\begin{eqnarray}
{\cal M}_{\rm KH}&=&1\times10^{10} (F/0.1)^{-7/2}(\rho_{\rm
ICM}/10^{-4}m_{\rm H})^{7/2}(\sigma_{\rm gal}/500{\rm km\, 
s^{-1}})^7
M_{\odot}
 \label{eq:mass_KH}
\end{eqnarray}
are difficult to be stripped by KHI. Since the time-scale of
KHI stripping is longer than that of ram-pressure stripping
\citep{Mori2000}, there remains a chance for galaxies that suffer
KHI to contribute to the BO effect temporarily before the
virialization of their cluster.

Here, we propose an alternative scenario for the BO effect, 
supposing
that the blueness of galaxies reflects large SFA. From
eq. (\ref{eq:mass_tid}), (\ref{eq:mass_ram}), and 
(\ref{eq:mass_KH}), we
find, when the mass of a galaxy is within the range of
\begin{eqnarray}
{\cal M}_{\rm KH} < {\cal M}_{\rm gal} < {\cal M}_{\rm tid}, 
\label{eq:result}
\end{eqnarray}
such less-luminous galaxies are affected neither by tidal 
interaction,
ram pressure, nor KHI. For such galaxies it is possible to retain
their ISM and SFA without the ISM being stripped, and hence 
contribute to BO 
galaxies.

Finally, we consider the evolution of clusters from the view of
their galaxies.  In un-virialized clusters, their galaxies are 
expected
to move relatively slow, and thus ${\cal M}_{\rm KH}$ becomes 
small, the
range of eq. (\ref{eq:result}) gets wide, and the fraction of blue
galaxies is high. On the contrary, in virialized clusters the 
opposite
situation is expected. Thus, the fraction of blue galaxies becomes
larger in clusters at higher redshift, according to their degree 
of
virialization.  Our scenario is consistent with the observational 
fact that
the fraction of blue galaxies is larger in irregular
(i.e. un-virialized) clusters whose $\sigma_{\rm gal}$ (velocity
dispersion among member galaxies) is smaller than those of regular
clusters \citep{Andreon1999, Margoniner2001}. The consistency of 
our
scenario with numerical simulations will be shown in Tajiri \& 
Yoshikawa
(in prep.).

\section{CONCLUSION}

We have investigated environmental effects on the SFA of 
star-forming cluster
galaxies within the magnitude range of $-22.5\le M_{\rm R}\le 
-19.6$ at
intermediate redshifts of $0.2<z<0.6$. The SFA is estimated from
EW([O{\sc ii}]). In less-luminous ($M_{\rm R}\gsim -20.7$) 
galaxies, we
firstly find that the mean SFA of cluster galaxies is smaller than 
that of
field galaxies. Secondly, we find that some of the cluster 
galaxies
have as large EW([O{\sc ii}]) as that of the actively star-forming 
field
galaxies. Such actively star-forming less-luminous cluster 
galaxies have
never been predicted before. To explain this new finding, we 
propose a
possible alternative scenario for the origin of Butcher-Oemler 
(BO)
galaxies. We propose that BO galaxies are stable against 
Kelvin-Helmholtz
instability and are not affected by tidal effects of a galaxy
cluster. Thus their ISM is not easily stripped and they keep on 
their
SFA. In younger clusters, such BO galaxies are more likely. As
clusters become virialized, however, member galaxies become more
accelerated, $\cal{M}_{\rm KH}$ increases, and fewer galaxies are 
stable against Kelvin-Helmholtz instability. Hence the 
fractions of blue galaxies decrease, which produces the BO effect.

  \acknowledgments

We thank the kindness of the two teams, Cl0024+1654 team (Czoske, 
O.,
Kneib, J.P., Soucail, G., Bridges,T., Mellier, Y., Cuillandre, 
J.C.),
and CNOC team (Balogh, M., Morris, S., Yee, H.K.C., Carlberg, R. 
\&
Ellingson, E.). They provided us with their data before 
publication. We
are also grateful to T.T.Takeuchi, K.Yoshikawa for their reading 
the
draft and giving a number of helpful suggestions, and to T.T.Ishii 
for
teaching how to make more appealing figures. We wish to express our
gratitude to an
anonymous referee for suggesting lots of useful improvements to our 
research and English. We made full use of 
the
NASA's Astrophysics Data System Abstract Service (ADS).

\newpage

\begin{table}[htbp]
\begin{center}
\caption{The results of Welch's {\it t}-test and Mann-Whitney's 
{\it
 U}-test for all galaxies in each magnitude bin (a left set) and 
those
 for galaxies with EW $<100$\AA~ (a right set)} \smallskip
\begin{tabular}{c|cccccc|cccccc}
\hline\hline
&\multicolumn{6}{|c|}{ALL}&\multicolumn{6}{|c}{BELOW 100 \AA}\\
$M_{R_i}$&$N_{\rm cl}$&$N_{\rm f}$&{\it
 t}&Sig.($t$)\tablenotemark{a}&{\it 
U}\tablenotemark{b}&Sig.($U$)\tablenotemark{c}&$n_{\rm
cl}$&$n_{\rm f}$&{\it t}&Sig.($t$)\tablenotemark{a}&{\it
U}\tablenotemark{b}&Sig.($U$)\tablenotemark{c}\\
\hline
-22.3&  ~~7 &  ~10 & -0.01 & 9.90e-01&~~31&6.96e-01& ~~7 &  ~10 & 
-0.01 &
9.90e-01&~~31&6.96e-01\\
-21.9&  ~23 &  ~30 & -0.50 & 6.17e-01&~318&6.28e-01& ~23 &  ~30 & 
-0.50 &
6.17e-01&~318&6.28e-01\\
-21.6&  ~50 &  ~67 & -2.05 & 4.29e-02&1310&4.43e-02& ~50 &  ~67 & 
-2.05 &
4.29e-02&1310&4.43e-02\\
-21.2&  106 &  ~65 & -3.89 & 1.80e-04&1935&1.55e-06& 106 &  ~65 & 
-3.89 &
1.80e-04&1935&1.55e-06\\
-20.8&  ~95 &  ~95 & -0.53 & 5.96e-01&3765&4.86e-02& ~94 &  ~95 & 
-1.60 &
1.11e-01&3670&3.45e-02\\
-20.5&  153 &  116 & -3.96 & 1.02e-04&6167&1.84e-05& 153 &  116 & 
-3.96 &
1.02e-04&6167&1.84e-05\\
-20.1&  130 &  ~96 & -2.30 & 2.22e-02&3873&1.10e-06& 129 &  ~96 & 
-3.89 &
1.37e-04&3777&5.68e-07\\
-19.7&  ~78 &  ~80 & -2.58 & 1.10e-02&1926&3.26e-05& ~76 &  ~79 & 
-4.36 &
2.40e-05&1767&9.76e-06\\
\hline
\end{tabular}
\end{center}
\tablenotetext{a}{Sig.($t$) means the significance level of each 
value of
 $t$. When Sig.($t$) is less than 0.05, the averages of two groups 
are
 significantly different.}  \tablenotetext{b}{$U$ means the rank 
sum of
 one group which has less samples.}  \tablenotetext{c}{Sig.($U$) 
means the
 significance level of each value of $U$. When Sig.($U$) is less 
than
 0.05, the representative statistical values of two groups are 
 significantly different.}
\end{table}

\clearpage

\begin{figure}[htbp]
\label{SFA_R}
\begin{center}
\leavevmode 
\caption{$R$-band magnitude and EW([O{\sc ii}]) of cluster 
galaxies
(Panel a) and those of field galaxies (Panel b).  In Panels (a) 
and (b),
black square indicates CNOC \citep{Balogh1999}, circle MORPHS
\citep{Dressler1999}, diamond Cl1358\citep{Fisher1998}, triangle 
AC103,
AC114, and AC118 \citep{Couch1998}, and white square Abell2390
\citep{Abraham1996}.}
\end{center}
\end{figure}

\begin{figure}[htbp]
\label{T_M}
\begin{center}
\leavevmode 
\caption{Mean EW([O{\sc ii}]) of cluster galaxies in each $R$-band
magnitude bin (Panel a) and that of field galaxies (Panel b). An 
error
bar designates the standard deviation of each value.}
\end{center}
\end{figure}

\newpage
\plotone{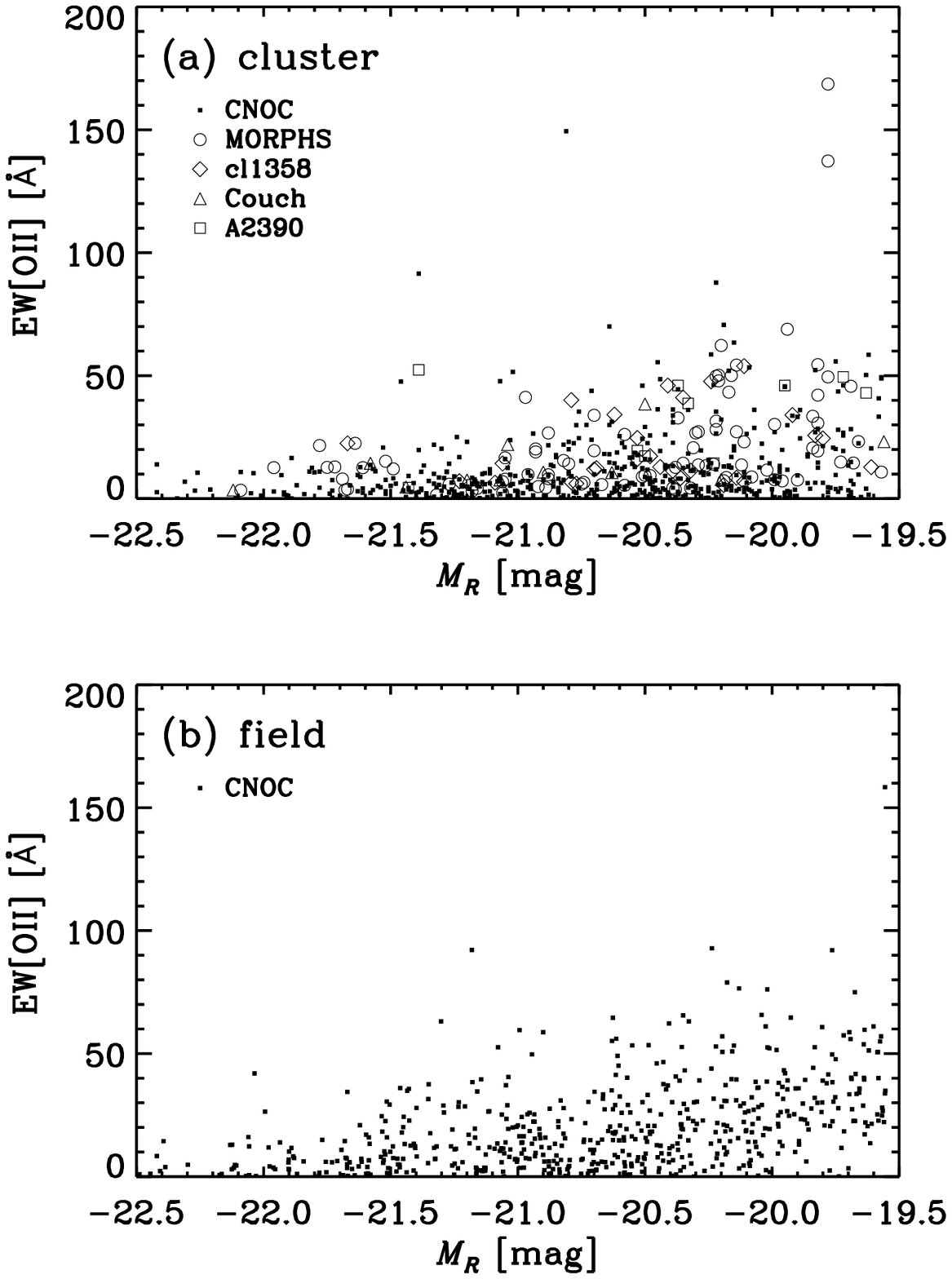}
\newpage
\plotone{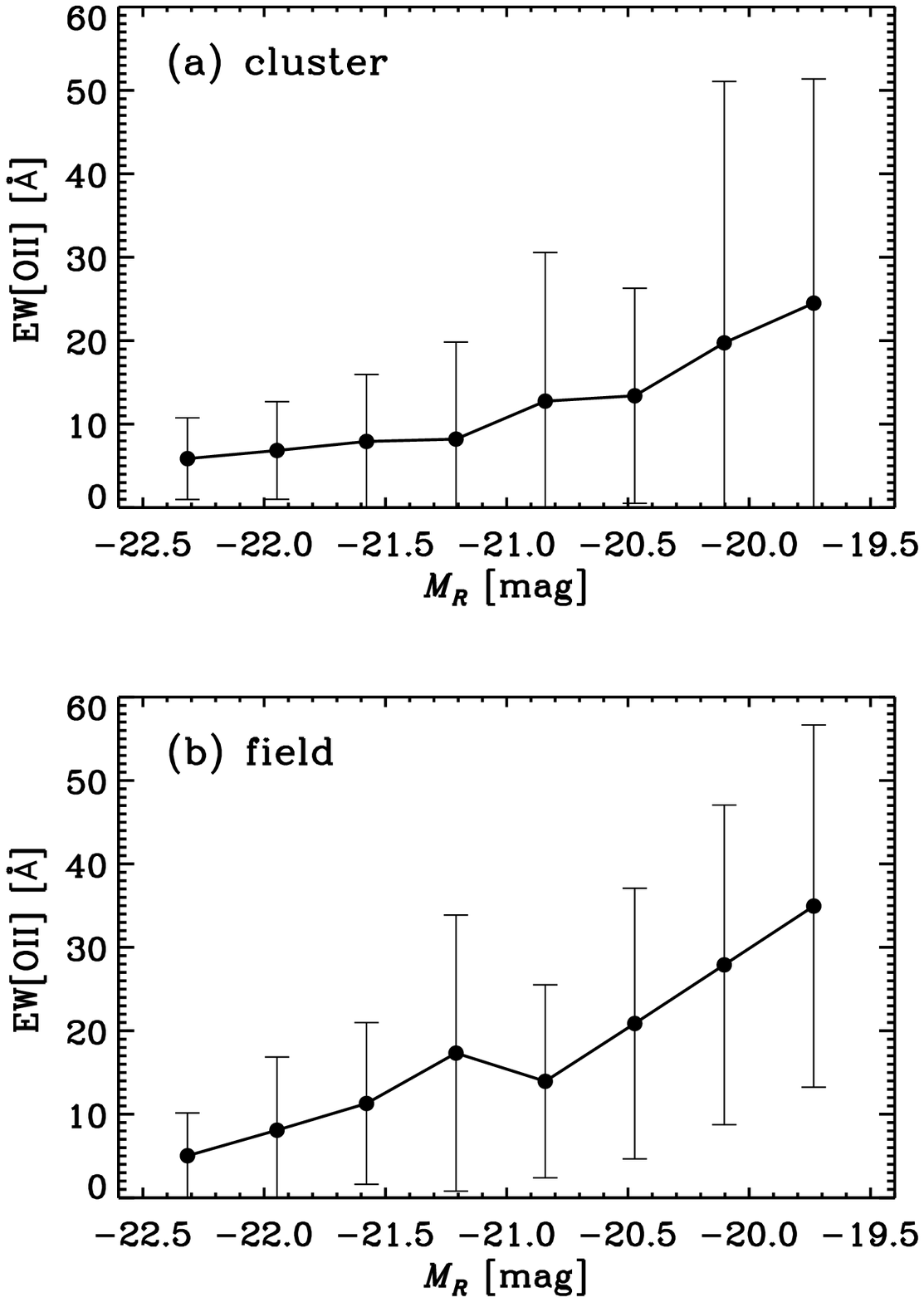}
\end{document}